\documentclass{iopart}
\usepackage{graphicx}
\usepackage{nicefrac}

\usepackage[T1]{fontenc}
\usepackage[latin1]{inputenc}

\usepackage{url}

\jl{31} 
\providecommand{\text}[1]{\textrm{#1}}

\begin{document}

\newcommand{\bskipdm}{\mskip -3.8\thickmuskip}
\newcommand{\bskiptm}{\mskip -3.0\thickmuskip}
\newcommand{\bskipsm}{\mskip -3.1\thickmuskip}
\newcommand{\bskipssm}{\mskip -3.1\thickmuskip}
\newcommand{\fskipdm}{\mskip 3.8\thickmuskip}
\newcommand{\fskiptm}{\mskip 3.0\thickmuskip}
\newcommand{\fskipsm}{\mskip 3.1\thickmuskip}
\newcommand{\fskipssm}{\mskip 3.1\thickmuskip}
\newcommand{\pint}{\mathop{\mathchoice{-\bskipdm\int}{-\bskiptm\int}{-\bskipsm\int}{-\bskipssm\int}}}
\newcommand{\abs}[1]{\left\vert#1\right\vert}
\newcommand{\ds}{\displaystyle}
\newcommand{\ts}{\textstyle}
\newcommand{\scs}{\scriptstyle}
\newcommand{\D}{\text{d}}
\newcommand{\I}{\mathrm{i}}
\newcommand{\EXP}[1]{\mathrm{e}^{#1}} 
\providecommand{\underset}[2]{\raisebox{-2.7mm}{$\stackrel{\ds #2}{\scs #1}$}}
\newcommand{\pabl}[3][\empty]{\frac{\partial^{#1} #2}{\partial {#3}^{#1}}}

\newcommand{\tpp}{\hat{t}}
\newcommand{\rpp}{\hat{r}}
\newcommand{\zpp}{\hat{z}}
\newcommand{\ppp}{\hat{\varphi}}
\newcommand{\tbb}{\bar{t}}
\newcommand{\rbb}{\bar{r}}
\newcommand{\zbb}{\bar{z}}
\newcommand{\pbb}{\bar{\varphi}}
\newcommand{\lbb}{\bar{\ell}}
\newcommand{\ttt}{\tilde{t}}
\newcommand{\rtt}{\tilde{r}}
\newcommand{\ztt}{\tilde{z}}
\newcommand{\ptt}{\tilde{\varphi}}
\newcommand{\ctt}{\tilde{c}}
\newcommand{\tvv}{{t^{\ast}}}
\newcommand{\tvvi}[1]{{t^{\ast}_{#1}}}
\newcommand{\xvv}{{x^{\ast}}}
\newcommand{\xvvi}[1]{{x^{\ast}_{#1}}}
\newcommand{\omvv}{{\omega^{\ast}}}
\newcommand{\kvv}{{k^{\ast}}}
\newcommand{\vvv}{{v^{\ast}}}
\newcommand{\lvv}{{{\ell}^{\ast}}}
\newcommand{\svv}{{s^{\ast}}}
\newcommand{\rvec}{\boldsymbol{r}}
\newcommand{\orig}{\boldsymbol{0}}
\renewcommand{\vec}[1]{\boldsymbol{#1}}
\def\clap#1{\hbox to 0pt {\hss #1\hss}}
\def\mathclap{\mathpalette \mathclapinternal}
\def\mathclapinternal#1#2{ %
\clap{$\mathsurround=0pt#1{#2}$}}

\title[A tale of two adventurers falling into a black hole]{Why ghosts don't touch: a tale of two adventurers falling one after another into a black hole}

\author{Klaus Kassner}
\address{Institut für Theoretische Physik, \\
  Otto-von-Guericke-Universität
  Magdeburg, Germany
}
\ead{Klaus.Kassner@ovgu.de}
\vspace*{2mm}
\begin{indented}
\item 26 August 2016
\end{indented}

\begin{abstract}
  The case for the utility of Kruskal-Szekeres coordinates in
  classroom made by Augousti et al.~in this journal
  (Eur.~J.~Phys.~33:1--11, 2012) is strength\-ened by extending their
  discussion beyond the event horizon of the black hole. Observations
  made by two adventurers following one another into a Schwarzschild
  black hole are examined  
  in terms of these nonsingular coordinates.  Two scenarios are
  considered, the first corresponding to one observer following the
  other closely, the second to a significant distance between the two
  of them, precluding the exist\-ence of a common inertial system.  In
  particular, the concepts of distance and temporal separation near the
  horizon and the redshift of the first infaller's image as seen by
  the second are investigated.  The results show that the notion of
  ``touching ghosts'' does not correspond to the local physics of two
  observers falling into a black hole. The story line is interesting
  enough and the mathematical details are sufficiently simple to use
  the example in a general relativity course, even at the
  undergraduate level.
\end{abstract}


\section{Introduction}
A discussion of the Schwarzschild spacetime is a mandatory part of any
full-fledged general relativity (GR) course.  Studying this metric,
we may derive predictions for the four classical tests of the theory
-- gravitational redshift, light deflection by gra\-vity, the
perihelion precession of planets, and the Shapiro delay. In addition,
the spherically symmetric vacuum metric gives access to some strong
field aspects, intro\-ducing fascinating new effects and exotic
physics, such as the phenomenon of an event horizon and the ensuing
existence of black holes.

Often, all of this is presented in terms of a single coordinate
system, nowadays called Schwarzschild coordinates. Unfortunately, this
preference gives rise to a certain amount of folklore about black
holes. Examples are the idea that an observer falling into a black
hole will take forever to reach the horizon or that a Schwarzschild black hole
cannot actually form, because the surface of the collapsing star that
should pro\-duce it takes infinite time to cross the incipient event
horizon, which then will never be present. Another element of folklore
holds that an observer, just before cross\-ing the horizon, will be able
to see the infinite future of the universe.  The first two state\-ments
are misconceptions based on mistaking a particular time coordinate,
the Schwarzschild time, for a substitute of Newton's absolute time,
the third is simply a falsehood \cite{grib09}.  An observer falling
towards the event horizon of a black hole does not approach the ``end
of time'' nor the ``end of the universe''. There are other coordinate
systems such as Gullstrand-Painlevé (GP) coordinates \cite{mueller10}
or Kruskal-Szekeres (KS) coordinates \cite{kruskal60, szekeres60},
sporting time variables that remain  finite as an infaller
crosses the event horizon. The GP time differs from the Schwarzschild
time only by a position dependent resynchronization transformation, whereas
the KS timelike coordinate is a more complicated mixture of the
Schwarzschild radial and temporal coordinates.

But what seems important then is that students are exposed to some of
these different coordinate systems in order not to misconceive the meaning
of Schwarzschild time, which may happen easily, if that is the only
time coordinate to which they ever are exposed in discussing 
non-rotating black holes. Therefore, it may be fruitful to look for
instances, where some of the other coordinate systems are either
necessary or at least beneficial.

In a recent article~\cite{augousti12}, KS
coordinates were argued to be a useful
pedagogical tool for discussing, in classroom, some non-trivial
results on observations near a black hole. As an example, the authors
of \cite{augousti12} propose to consider the possibilities for (radio)
signal exchange and optical monitoring between two  observers
falling successively into a black hole.
Interestingly, they do not exploit these coordinates to their full
power -- they avoid discussing observations \emph{after} either of the
two adventurers has crossed the event horizon. While different
philosophical attitudes may be taken towards the relevance of
predictions made by GR about what happens inside
the horizon -- those events remain invisible to the outside world and
inaccessible to any non-suicidal observer -- GR is a deterministic
theory and offers precise statements about this spacetime region.
It is not a priori unscientific to consider and discuss these
predictions.

Therefore, amplifying on the assertion of pedagogical value of
Kruskal-Szekeres coordinates, I would like to complement the paper by
Augousti \emph{et al.}~\cite{augousti12} with an investigation about
what happens beyond, and in particular on, the event horizon,
especially in view of the authors' own suggestion
that there is grounds for further discussion.  Moreover, it seems to
me that a few statements made in \cite{augousti12} are not borne
out by closer inspection.  In particular, I will argue that the idea
of the second infalling observer (Bob) perceiving to touch a ghostly
image of the first (Alice) on approaching the horizon is incorrect
and, in fact, incompatible with the equivalence principle. Also, Bob
will not find Alice pass the horizon at the same \emph{instant} as he himself
crosses it. While he will \emph{see} her traversing the horizon
the moment when he himself crosses it, he will see her at a distance
(despite their Schwarzschild radial coordinate being the same) and he
will conclude her to have passed the horizon before him, because the
light signal took time to travel from her position to his.

Confusion about the interpretation of calculational results may arise
from insufficient care in distinguishing between global and local
coordinates and lack of appreciation of the fact that having similar
or even the same $(r,\vartheta,\varphi)$-coordinates does not
necessarily mean physical closeness when $r=r_S$, the Schwarzschild
ra\-dius. At that radius, Schwarzschild coordinates become singular, and
the ratio be\-tween the radial proper length element of a coordinate stationary
observer and $\D r$ tends to infinity. Neither Alice nor Bob have a
finite (Schwarzschild) time coordinate at the horizon. Discussing
position differences and temporal separation near the horizon in terms of
Schwarzschild coordinates is challenging and not really recommendable.

Kruskal-Szekeres coordinates are clearly better, they show immediately
that the horizon crossings of the two observers correspond to two
different events. These are separated by a finite (null)
spacetime interval. How this is locally perceived in terms of space
and time intervals, is best investigated by a transformation to local
inertial coordinates of either observer. If the two observers start
their journey sufficiently closely, there may even exist a common
local inertial system near the horizon, in which the discussion
becomes very simple. Beyond that, inferences have to be made using
global coordinates among which KS coordinates belong to the more
useful ones.  Another advantageous choice would be GP coordinates, in
which the radial and time coordinates are more easily interpretable
and which are also regular at the horizon, but where the graphical
representation of light cones is not as simple as in KS
diagrams.

The remainder of the paper is organized as follows. In
section~\ref{sec:coord_rep}, the Schwarzschild metric is given both in
its standard form and in KS coordinates. The latter re\-presentation
is needed for the calculation of KS diagrams that will be used in
section~\ref{sec:ks_diag} to visualize the free-fall journeys. This
section is the one with the highest pedagogical utility. In
section~\ref{sec:local_desc}, some simple analytical calculations are
given that clarify what the situation will look like locally. This
kind of material might be used in supplementary exercises of a GR
course. Some conclusions summarize our results.  An appendix shows how
to obtain the redshift of an infalling observer, seen by another
infaller, when the latter reaches the event horizon. This material
could be offered to students as a homework calculation (for bonus
points).

\section{Coordinate representations}
\label{sec:coord_rep}

To fix the notation, we introduce the Schwarzschild metric in the form:
\begin{equation}
\hspace*{-1cm}\eqalign{\D s^2 &= \left(1 - \frac{r_S}{r}\right)\,c^2 \D t^2 -\left(1-\frac{r_S}{r}\right)^{-1}\, \D r^2
- r^2 \left(\D \vartheta^2 + \sin^2 \vartheta \,\D \varphi^2\right) \cr
&= g_{tt}\, c^2 \D t^2 + g_{rr} \, \D r^2+ g_{\vartheta \vartheta}\,\D \vartheta^2 + g_{\varphi\varphi}\,\D \varphi^2}
\end{equation}
where 
\begin{equation}
r_S = \frac{2GM}{c^2}
\end{equation} 
is the Schwarzschild radius. KS coordinates $(v,u,\vartheta,\varphi)$ are obtained by the coor\-di\-nate transformation  \cite{mueller10} 
\begin{equation}
\eqalign{\hspace*{-2cm}u &= \sqrt{1-\frac{r}{r_S}}\, \EXP{r/2r_S} \sinh\frac{ct}{2r_S}\>, \quad 
v =  \sqrt{1-\frac{r}{r_S}}\, \EXP{r/2r_S}\cosh\frac{ct}{2r_S}\>,  \quad 0<r<r_S\>,\\
\hspace*{-2cm}u &=\sqrt{\frac{r}{r_S}-1} \,\EXP{r/2r_S} \cosh\frac{ct}{2r_S}\>, \quad v 
=  \sqrt{\frac{r}{r_S}-1}\, \EXP{r/2r_S}\sinh\frac{ct}{2r_S}\>,  \quad r\ge r_S\>.}
\end{equation}
In these coordinates, the line element is given by
\begin{equation}
 \D s^2 =\frac{4 r_S^3}{r} \, \EXP{-r/r_S}\left(\D v^2-\D u^2\right) - r^2 \left(\D \vartheta^2 + \sin^2 \vartheta \,\D \varphi^2\right)\>.
\label{eq:ks_line_el}
\end{equation}
The Jacobian of the relevant part of the transformation is \cite{augousti12}
\begin{equation}
\frac{\partial{(v,u)}}{\partial{(c t,r)}} = \left[\begin{array}{cc} \frac{u}{2 r_S} &  \frac{v}{2 r_S g_{tt}(r)} \\
 \frac{v}{2 r_S} &  \frac{u}{2 r_S g_{tt}(r)} 
\label{eq:jacob}
  \end{array}\right]
\end{equation}
and its  determinant reads
\begin{equation}
D\equiv \left|\frac{\partial{(v,u)}}{\partial{(c t,r)}}\right| = \frac{r}{4 r_S^3}\,\EXP{r/r_S}\>,
\end{equation}
which is just the inverse of the prefactor of $ \D v^2-\D u^2$ in (\ref{eq:ks_line_el}). Other useful formulas are \cite{mueller10}
\begin{equation}
\eqalign{\hspace*{-1cm}\left(\frac{r}{r_S}-1\right) \EXP{r/r_S} =u^2-v^2\>,\qquad
  &r=r_S\left[\mathcal{W}\left(\frac{u^2-v^2}{\text{e}}\right)+1\right]\>,\cr
 \hspace*{-1cm} c t= 2 r_S \,\text{arctanh} \frac{v}{u} \>,\quad r> r_S \>,\qquad &c t= 2 r_S\, \text{arctanh} \frac{u}{v} \>,\quad r< r_S \>,}
\label{eq:useful_formulas}
\end{equation}
where $\mathcal{W}$ is Lambert's W function \cite{corless96}. The first of these
equalities shows that $r=r_S$ corresponds to $\abs{v}=\abs{u}$ and
that other constant $r$ values are described by hyperbolas in the $uv$
plane.
From the second line, we note that constant time $t$ corresponds to a
fixed ratio $v/u$, i.e., a straight line emanating from the origin in
the $uv$ plane. The event horizon corresponds to $r=r_S$ with $u=v
>0$, so it has the Schwarzschild time coordinate $t=\infty$, i.e.,
that time coordinate is rather useless for its description.

\section{Kruskal-Szekeres diagrams}
\label{sec:ks_diag}

Having prepared the mathematical details, let us return to the tale of
Alice and Bob.  To avoid a situation in which our adventurers would be
torn apart by tidal forces before reaching the event horizon, we
assume that the black hole under consideration is supermassive,
comprising a few hundred million solar masses at least. Then surface
gravity at the horizon is weak and local inertial systems there are
neither extremely small nor very short-lived.

Consider Alice freely falling into the black hole, followed by Bob.
We may then distinguish two situations. Either Bob follows Alice so
closely that both may be considered being in a single local inertial
system (for some time) or he starts his journey so much later than
Alice that this assumption is not satisfied anywhere near the horizon.
Let us call the first situation scenario I, the second scenario II.
Figure~\ref{fig:ks} depicts these cases in KS diagrams, the left panel
visualizing scenario I, the right one scenario II \footnote{Alice's
  and Bob's trajectories in figure 1 were calculated by numerical
  integration of (\ref{eq:vdot_2}) and (\ref{eq:udot_2}).}.  Alice's
trajectory is the same in both images, and all the trajectories start
at $r=r_0$ at zero velocity, with only the departure times of Bob
being different in the two cases. With this arrangement, we can
immediately tell the difference in proper times of the two adventurers
on crossing the horizon, if they synchronized their clocks before the
start: since they fall on identical spatial trajectories they both
need precisely the same proper time interval from the beginning of
their fall to the horizon.  Therefore, at the horizon their clocks
will differ by the proper time difference $\tau_B-\tau_A$ of their
moments of departure.

\newlength{\textwidthpart}
\setlength{\textwidthpart}{0.425\textwidth}

\begin{figure}[htb!]
\rule{0pt}{0pt}\hfill
\includegraphics[width=\textwidthpart]{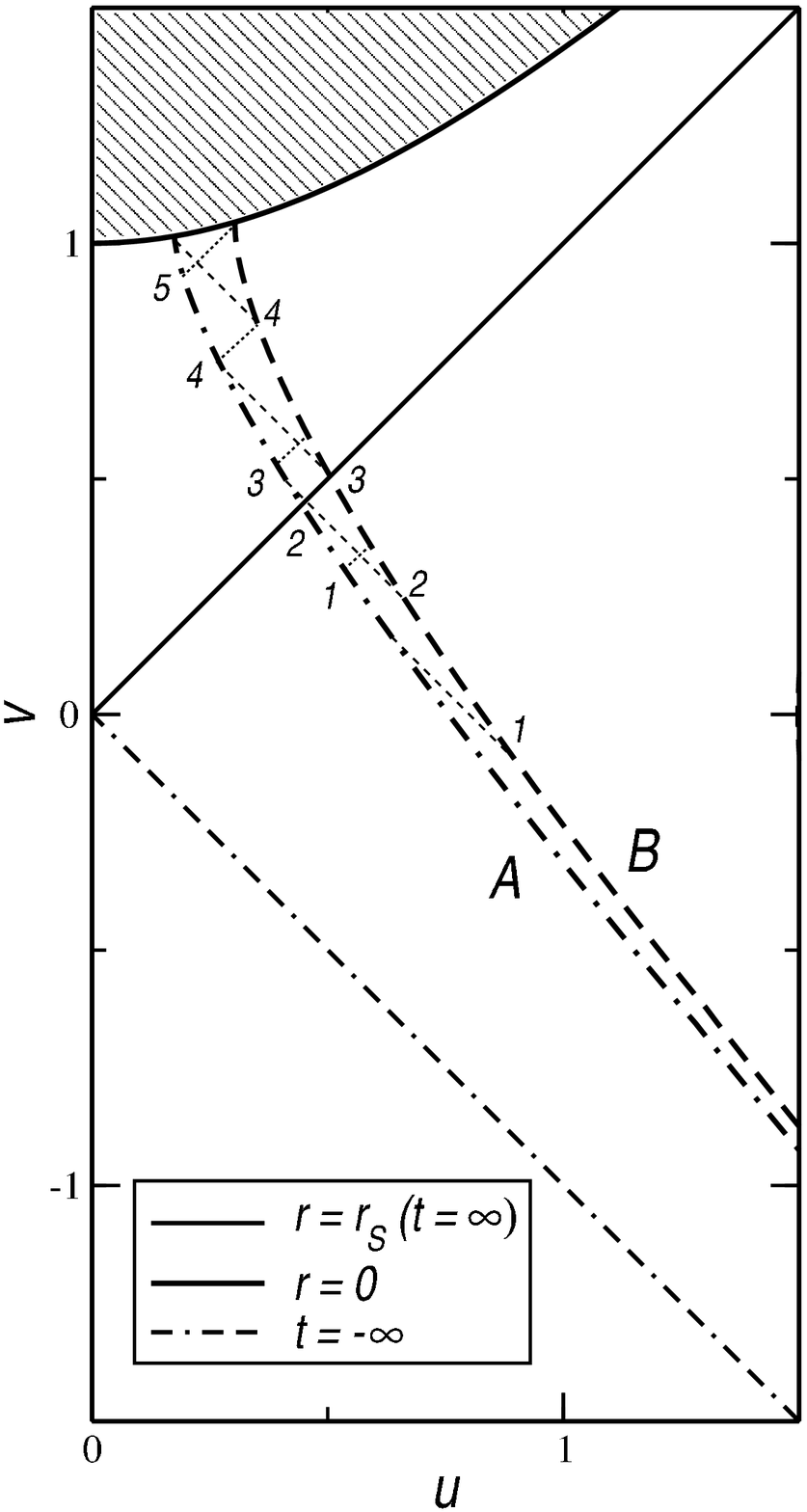}
\hspace*{0.5cm}\includegraphics[width=\textwidthpart]{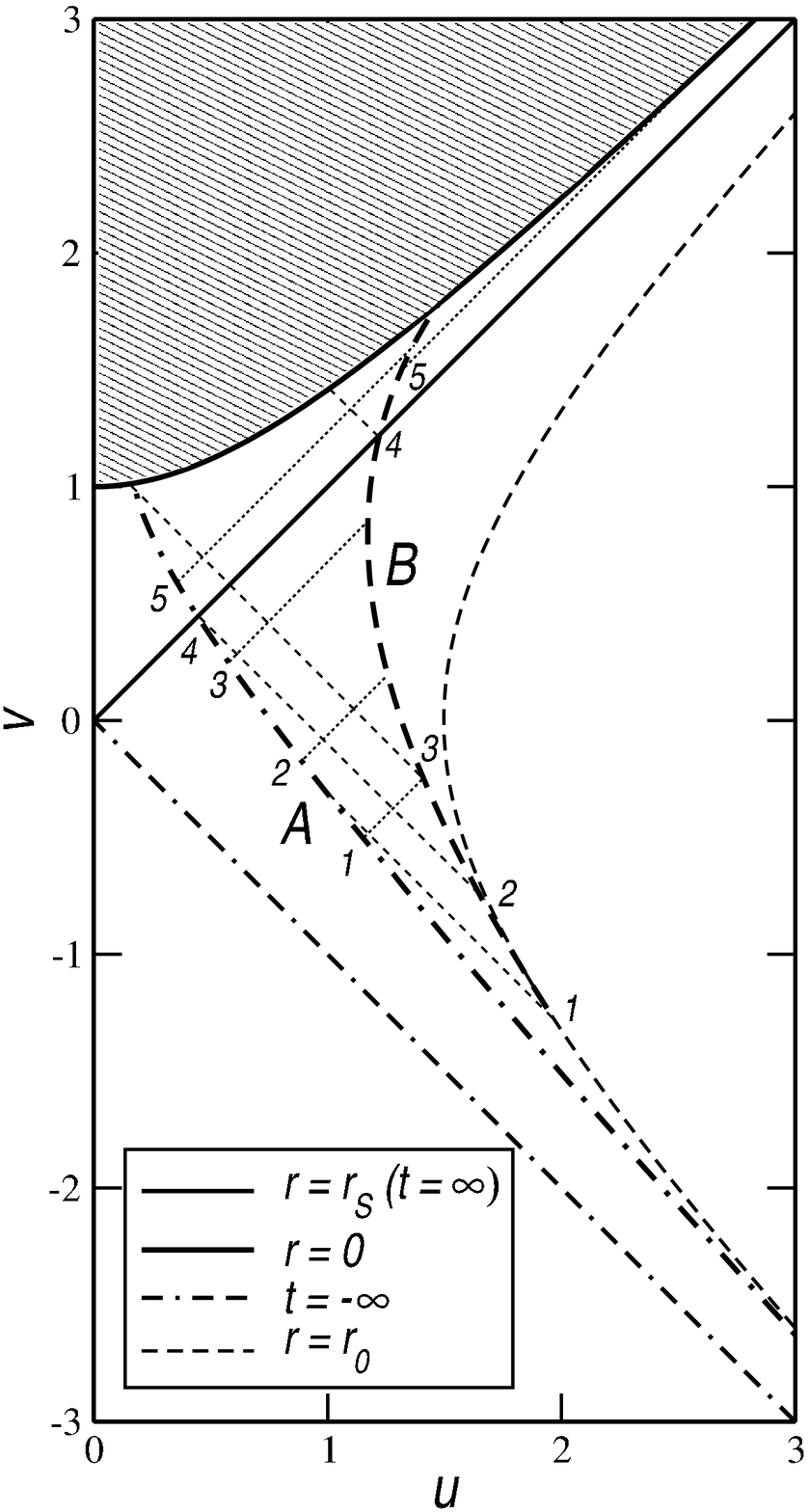}
\hfill\rule{0pt}{0pt}
\caption{
Left: Alice ($A$) falls into the black hole with Bob ($B$) following
immediately.  Right: Alice on the same trajectory, but now Bob starts
his journey later. Parameters used in the numerical 
calculation of the graphs are $r_S=2$,
  $r_0=3$; the starting time of Alice is $t_A=-7$, that of Bob
  $t_{B1}=-6.5$ and $t_{B2}=-3$, respectively.
\label{fig:ks}
}
\end{figure}
\newcommand{\sfsl}[1]{\textsf{\textit{#1}}}

In scenario I, obviously nothing particular will happen at the horizon
from the point of view of our observers. They should be able to
continually exchange messages without noticing their traversing the
horizon. This is granted by the equivalence principle which says that
locally their physics is that of special relativity. There is
no fundamental reason for the equivalence principle to fail at the
horizon \footnote{We stay within classical physics in this paper, so we
  do not have to discuss recent attacks on the equivalence principle
  based on the firewall conjecture \protect\cite{almheiri13}.}.
Therefore, in particular, there will be no ghostly image of Alice
approaching Bob, as special relativity does not envisage such an
effect. The only way to escape from this conclusion would be to deny the
possibility of scenario I. However, the size of a local inertial
system near the horizon can be made as large as we wish, as long as we
may consider arbitrarily massive black holes.

Moreover, the figure demonstrates that signal exchange is not
disturbed by the presence of the horizon. The numbers \sfsl{1} to
\sfsl{5} in the left panel along the trajectory $A$ and \sfsl{1} to
\sfsl{4} along $B$ refer to a few light or radio signals sent from
Alice to Bob and vice versa. Bob's signal \sfsl{2} reaches Alice as
she crosses the horizon, whereupon she sends her signal \sfsl{2},
reaching Bob at point \sfsl{3}, precisely when he is about to cross
the horizon.  Signal \sfsl{3} from Bob catches up with Alice before
she hits the singularity at $r=0$ ($v^2-u^2=1$), and she still has
time to answer with signal \sfsl{4}. KS coordinates are nonsingular
outside $r=0$, so if there is a null trajectory connecting Alice's and
Bob's world lines, light can travel along it. 

The authors of \cite{augousti12} state that no signal can be emitted
at the horizon (because there could be no ``trapped'' signal). This
seems to be based on a very common mis\-conception about the nature of
the event horizon, which is to visualize the horizon as being
\emph{static}. It is well-known and may be directly read off a KS
diagram that the ho\-rizon is a null hypersurface. Hence, for
any inertial (i.e.~freely falling) ob\-ser\-ver close to it the horizon
moves (outward) at the speed of light!  This fact becomes particularly
transparent in the river model of a spherical black hole
\cite{hamilton08}, where space is streaming inward, toward the center
of symmetry and reaches the speed of light at $r=r_S$. Thus,
in order to stay put there, one would have to travel outward as fast
as a photon, which the horizon does (and can do, not being a material
object).

It is then quite easy to understand how a signal from Alice sent the
moment she crosses the horizon will reach Bob the very moment he does
so, too, in spite of the fact that this happens \emph{later}. The solution is
similar to that of the puzzle whether Alice when falling through with
her feet first will cease to see them for a moment. Clearly, she will
have no trouble seeing her feet all the time, because the time light
from her feet crossing the horizon takes to her eyes is precisely the
time the horizon itself takes to travel to her eyes as well. The
relative speed between observer and horizon is $c$ at the moment of
passage. Bob will receive Alice's signal when he reaches the horizon,
because the horizon is moving towards him. If he fired his thrusters
to get away, he might outrun the horizon, but then he would also
successfully run away from Alice's signal. And of course, he will
never have the impression that Alice is going to collide with him. Such
an idea would only arise from a misinterpretation of the meaning of
the coordinate $r$ as an absolute measure of distance. Her (proper)
distance from Bob is in fact continually increasing as will be
explained below.

These qualitative statements may be made more precise using the KS
diagram. Since KS coordinates are not singular at the horizon, the
fact that Alice and Bob have different coordinates $(u_A,v_A)$ and
$(u_B,v_B)$ on passing the horizon means that these crossings are
\emph{different events}. But different events cannot correspond to
both the same time and the same place. In fact, since two of the four
coordinates ($\vartheta=\text{const.}$ and $\varphi=\text{const.}$)
are the same for both travelers, the fact that $(u_A,v_A)$ and
$(u_B,v_B)$ have \emph{null} separation implies that Alice and Bob
will perceive their passage of the horizon to happen \emph{neither} at
the same place \emph{nor} at the same time \footnote{$ds^2=0$ and
  equality of \emph{three} (nonsingular) coordinates implies equality
  of all four coordinates for both events. But then the events would
  not correspond to different spacetime points. So if one pair of
  coordinates are different, there must be another pair that are
  different, too.}. Moreover, we know that two different events may be
considered happening at the \emph{same place} only if they can be
connected by a \emph{timelike} world line (which is not the case here,
the connection is null). Hence, Bob will not see a ghostlike image of
Alice approaching him. Rather, he will see her at constant distance as
long as they may be considered sharing a common inertial system and
she will drift away as soon as tidal forces become noticeable.

An interesting question is whether Alice and Bob will agree on who hits
the singularity first. Alice would receive Bob's signal \sfsl{4} the
moment she hits the singularity (had she not been spaghettified
before... \cite{wiki_spaghettification}) so Bob would not yet have hit
it, because he could not pass his own radio signal. However, this
argument is based on the prejudice that the singularity corresponds to
a single event, which it clearly does not. Bob receives signal
\sfsl{5} from Alice when hitting the singularity, so he might make up
a similar argument to show that Alice will survive him for a few
moments. But this is not conclusive, because Alice's signal was sent
opposite to her direction of motion.  Near the singularity, Alice and
Bob will not share a common local inertial system, so the answer is
not unambiguous, given the fact that the singularity is
spacelike \footnote{This can be seen from the KS diagram -- the
  magnitude of the slope of the curve $v^2-u^2=1$ is smaller than 1
  everywhere.}. In terms of the timelike variable $v$, it is clearly
Alice who hits the singularity first.

Let us now discuss scenario II.  If there is no ghostly image in
scenario I, it is hard to believe that such an image should appear in
scenario II, where Bob follows Alice at a larger distance. Indeed,
there never can be a situation of \emph{touching} ghosts for our two
adventurers, as long as they remain far from each other. But if they
approach one another more closely than the typical diameter of a local
inertial system, we are back to scenario I, in which the equivalence
principle forbids such a phenomenon and in which residual tidal forces
will tend to make them drift apart rather than draw near. All that
\emph{might} be possible in scenario II, then, is some \emph{optical
  illusion}.

Therefore, let us turn to the issue whether Alice's image as perceived
by Bob, if not appearing to touch him in the guise of a ghost, may at least
have \emph{ghostly features}. If Bob starts his journey sufficiently
much later than Alice, i.e., if he remains long enough at their
stationary\footnote{By \emph{stationary} we always mean being at rest
  in the $(t,r,\vartheta,\varphi)$ system, i.e., $\D r = \D\vartheta =
  \D\varphi=0$, that is, coordinate stationary in Schwarzschild
  coordinates. Coordinate stationary observers in KS coordinates would
  be moving observers in Schwarzschild coordinates.}  ``mother
station'' \cite{augousti12}, he will of course see her slowing down
and her image redshift as she approaches the horizon.  This is the
standard sce\-nario seen by a stationary distant observer.  As Bob then
falls towards the horizon him\-self, he will pick up speed, which should
compensate for part of Alice's redshift. A full compensation seems
however unlikely, given the fact that Alice is always in a more
strongly accelerating field than Bob. Rather, Bob's motion should lead
to Alice's image redshifting not quite as fast as if he had stayed
behind. This is shown to be true in the appendix.  If Bob actually
follows her through the horizon, her image will in fact never seem to approach a complete freeze.

A quantity that can be easily calculated is Alice's
redshift as perceived by Bob when he passes the horizon. The
calculation gives a finite result and is exhibited
in the appendix. Together with
figure~\ref{fig:ks} it implies that Alice's image is, not
surprisingly, the more redshifted on Bob's reaching the horizon, the
later Bob starts his journey. Therefore, if he follows her too late,
i.e., when optical frequencies emitted by her have already
shifted outside the visible spectrum, he will not see her again and
her ghost less than ever.

What does he see, if either he follows early enough for her image to
be still visible or if he uses appropriate infrared vision
gear\footnote{The time scales on which this kind of equipment may be
  useful are pretty short -- in the second range for a black hole with
  a million solar masses. In order to allow for  interesting optical effects
  in the late-time scenario discussed here, either a black hole in the
  range above a billion solar masses is needed or we must consider a
  few milliseconds a long time.}?  First, aside from redshifting and
acquiring a slow-motion aspect, Alice's image will fade quickly. Light
emitted by her body, will, as far as it is sent slightly outside or inside the
horizon, diverge away from the latter, escaping to infinity or falling
towards the singularity, so the electromagnetic energy deposited near
the horizon will diminish fast, and her image will lose intensity
equally fast. Second, while in the local scenario Alice will seem to
move away from Bob slowly due to small tidal forces as the
approximation of a common inertial system gets worse, Alice will \emph{not}
seem to move away to \emph{arbitrarily large} distances in scenario II, as we
make the time delay between hers and Bob's journey larger.  The
horizon has a finite area and her image hovering there will take a
finite fraction of that area. Since we are limited, in our simple
approach, to the discussion of temporal and radial aspects, we cannot
describe here the distortions and size effects that Bob will see on
passing the horizon some substantial time after Alice. The result of
such a discussion would be that there is a maximum distance at which
(a strongly redshifted) Alice will appear to Bob on entering the
region interior to the event horizon, dependent on the size of the
black hole. She will not seem to have receded to infinity, regardless of how
much time he lets pass before following her \cite{hamilton16}.

Both travelers will agree in both scenarios that it is Alice who
passes first and Bob who passes last. They never have the feeling
of passing at the same time.

The main difference between the two scenarios is that in scenario II
Bob cannot send signals to Alice anymore, once he has passed the
horizon. His signal \sfsl{4} in the right panel of figure \ref{fig:ks}
obviously can never reach Alice, nor can signals sent later. 
Note that the last signal Alice can
receive from Bob (signal \sfsl{3}) was sent while he was still
outside the horizon. Alice will never know whether Bob tried to escape
the black hole after sending that signal. On the other hand, Bob will
still continue to receive signals from Alice until he hits the
singularity. Again the signal sent by Alice while crossing the horizon
(signal~\sfsl{4}) will reach Bob only the moment he passes the
horizon, too, because it cannot outrun the outward-moving horizon.

\section{Local description of scenario I}
\label{sec:local_desc}

It is possible to calculate the proper distance and proper time
interval of a freely falling observer at the horizon, allowing Alice
and Bob to explicitly derive the spatial and temporal distance between
the two events of their horizon crossings in the case of scenario I.
To this end, we would like to formulate these quantities in terms of
KS coordinates. There are (at least) two ways to achieve this. The
pedestrian approach consists in first expressing the velocity of the
infaller by the proper time element $\D t_s$ and radial proper length
element $ \D\ell_s$ of a stationary observer next to him.  (This works
only outside the horizon.)
Then a local Lorentz transformation
to the frame of the falling observer whose velocity is
$v_f=\D\ell_s/\D t_s$ produces the proper time and length elements of
the infaller in terms of $\D t$ and $\D r$, a result that may be
straightforwardly transformed to KS coordinates.

The second, somewhat faster approach avoids any explicit Lorentz
transformation. We simply determine the local coordinate
transformation that takes the KS metric to Minkowski form in the frame
of the freely falling observer. The results of both approaches are the
same, of course.

Let us name the proper time element of
the infaller $\D \tau$ and her proper (radial) length element $
\D\ell$, then there must be a relationship (assuming
$\D\vartheta=\D\varphi=0$)
\begin{equation}
\eqalign{
  \D v &=\alpha \D \tau + \beta \D\ell \>, \cr
  \D u &=\gamma \D \tau + \delta \D\ell \>,}
\label{eq:gen_rel_elements}
\end{equation}
and we must have $\alpha/\gamma= \dot v/\dot u$ (where $\dot v = \D
v/\D \tau$, $\dot u = \D u/\D \tau$), because the proper time is a
local coordinate that is tangent to the world line of the observer.
From the line element
\begin{equation}
 \D s^2 =D^{-1} \left(\D v^2-\D u^2\right) 
\label{eq:ks_line_sh}
\end{equation}
we gather that
\begin{equation}
\dot v^2-\dot u^2 = D c^2\>.
\label{eq:vdot2_udot2}
\end{equation}
To determine the coefficients of the transformation we require
\begin{equation}
  \eqalign{\hspace*{-2cm}D^{-1} \left(\D v^2-\D u^2\right) &= {D^{-1} 
\left[\left(\alpha^2-\gamma^2\right) \D \tau^2 
        + 2\left(\alpha \beta-\gamma\delta\right) \D \tau \D\ell+ \left(\beta^2-\delta^2\right)
        \D\ell^2 \right] } \cr
    &\stackrel{!}{=} c^2 \D \tau^2 - \D\ell^2\>.}
\end{equation}
This gives us three equations
\begin{equation}
 \eqalign{ \alpha^2-\gamma^2 &= D c^2 \>,\cr 
\alpha\beta - \gamma\delta &= 0\>,  \cr
\delta^2-\beta^2 &= D\>.
}
\end{equation}
With $\alpha = \gamma \dot v/\dot u$, the second equation yields 
$\delta = \beta \dot v/\dot u$. The remaining ones give
\begin{equation}
  \eqalign{\gamma^2\left(\frac{\dot v^2}{\dot u^2} -1\right )
    &= D c^2  \quad\Rightarrow\quad 
    \gamma^2\left(\dot v^2-\dot u^2\right) \underset{(\ref{eq:vdot2_udot2})}{=}
    D c^2 \gamma^2 = D c^2 \dot u^2\>, \cr
    \beta^2 \left(\frac{\dot v^2}{\dot u^2} -1\right ) 
    &= D  \phantom{c^2}   \quad\Rightarrow\quad 
    \beta^2\left(\dot v^2-\dot u^2\right) =D c^2 \beta^2 = D  \dot u^2 \>.
  }
\end{equation}
Assuming $\dot v>0$, we finally obtain
\begin{equation}
\alpha = \dot v\>, \quad\beta = \frac{\dot u}{c}\>, \quad \gamma =\dot u\>, \quad\delta =\frac{\dot v}{c}\>,
\end{equation}
and solving (\ref{eq:gen_rel_elements}) for $\D\tau$ and $\D\ell$, we have
\begin{equation}
\eqalign{\D\tau &= \frac{1}{D c^2} \left(\dot v \D v - \dot u \D u\right) \>, \cr
\D\ell &= \frac{1}{D c} \left(\dot v \D u - \dot u \D v\right)\>.
\label{eq:proper_fall}
}
\end{equation}
Since we did not use the fact anywhere that our observer is freely
falling -- geodesic equations were not invoked -- this form of the
result is true for observers in arbitrary motion. To apply it to the
case of Alice and Bob, we need their four-velocity com\-ponents $\dot v$
and $\dot u$ in the actual free-fall situation. Instead of trying to
solve for them in KS coordinates directly, we use the equations of
motion from \cite{augousti12} in Schwarzschild coordinates and
transform to KS coordinates. For an observer starting at $r_0$ with
zero velocity, the four-velocity components in the $(t,r)$ system are
\begin{equation}
\eqalign{ V^t = c \dot t \equiv c \frac{\D t}{\D \tau} &= c \frac{\ds\sqrt{1-\frac{r_S}{r_0}}}{\ds 1-\frac{r_S}{r}}\>, \cr
 V^r = \dot r \equiv \frac{\D r}{\D \tau} &= -c  \sqrt{\frac{r_S}{r}-\frac{r_S}{r_0}} \>.
}
\label{eq:fourvel_schwarz}
\end{equation}
We then have, using (\ref{eq:jacob})
\begin{equation}
  \eqalign{\hspace*{-10mm}V^v=\dot v &=\frac{\partial v}{\partial t} \dot t
    +\frac{\partial v}{\partial r}  \dot r 
    = \frac{c}{2 r_S} \frac{1}{\ds 1-\frac{r_S}{r}}\left(u\sqrt{1-\frac{r_S}{r_0}}
      -v \sqrt{\frac{r_S}{r}-\frac{r_S}{r_0}}\right)\>,\cr
   \hspace*{-10mm}V^u= \dot u &=\frac{\partial u}{\partial t} \dot t+\frac{\partial u}{\partial r} \dot r 
    = \frac{c}{2 r_S} \frac{1}{\ds 1-\frac{r_S}{r}}\left(v\sqrt{1-\frac{r_S}{r_0}}-u
      \sqrt{\frac{r_S}{r}-\frac{r_S}{r_0}}\right)\>.
  }
\label{eq:vdot_udot}
\end{equation}
Next, we wish to convince ourselves that $\dot v$ and $\dot u$ do not
turn singular at the horizon $r=r_S$, in spite of the denominator
$1-\frac{r_S}{r}$. A simple trick permitting to  demonstrate this analytically  for
$\dot v$ and giving a manifestly nonsingular formula is to multiply
the numerator and denominator by $u\sqrt{1-\frac{r_S}{r_0}}+v
\sqrt{\frac{r_S}{r}-\frac{r_S}{r_0}}$.  This produces
\begin{equation}
\hspace*{-2cm}  \eqalign{\dot v &=  \frac{c}{2 r_S} 
    \frac{1}{1-\frac{r_S}{r}}\frac{u^2\left(1-\frac{r_S}{r_0}\right) 
      -v^2\left(\frac{r_S}{r}-\frac{r_S}{r_0}\right) }{u\sqrt{1-\frac{r_S}{r_0}}  
      + v\sqrt{\frac{r_S}{r}-\frac{r_S}{r_0}} }\cr
    &=  \frac{c}{2 r_S}\frac{\left(u^2-v^2\right)\left(1-\frac{r_S}{r_0}\right) 
      +v^2\left(1-\frac{r_S}{r}\right)}{\left(1-\frac{r_S}{r}\right)
      \left(u\sqrt{1-\frac{r_S}{r_0}}  
        + v\sqrt{\frac{r_S}{r}-\frac{r_S}{r_0}}\right)}
    =  \frac{c}{2 r_S}\frac{v^2+\frac{r}{r_S}\,\EXP{r/r_S}
\left(1-\frac{r_S}{r_0}\right) }{u\sqrt{1-\frac{r_S}{r_0}}  
      + v\sqrt{\frac{r_S}{r}-\frac{r_S}{r_0}} }
  }
\label{eq:vdot_2}
\end{equation}
with the regular limit 
\begin{equation}
  \lim_{r\to r_S} \dot v =   \frac{c}{2 r_S} \frac{v^2+\text{e} 
    \left(1-\frac{r_S}{r_0}\right)}{(u+v) \sqrt{1-\frac{r_S}{r_0}} }\>,
\end{equation}
which may be simplified a bit more by using that $u=v$ on the horizon.
An analogous calculation for $\dot u $ yields:
\begin{equation}
  \dot u =  \frac{c}{2 r_S}\frac{u^2-\frac{r}{r_S}\,\EXP{r/r_S}
    \left(1-\frac{r_S}{r_0}\right) }{v\sqrt{1-\frac{r_S}{r_0}}  
    + u\sqrt{\frac{r_S}{r}-\frac{r_S}{r_0}} }\>.
\label{eq:udot_2}
\end{equation}
We then obtain for the proper length element of a freely falling
observer, using (\ref{eq:proper_fall}),
\begin{equation}
\hspace*{-2cm}\D \ell = 
    2 r_S\left\{\rule{0pt}{7mm}\right. 
    \frac{\frac{r_S}{r}\,\EXP{-r/r_S} v^2+1-\frac{r_S}{r_0} }{u
      \sqrt{1-\frac{r_S}{r_0}}  
      + v\sqrt{\frac{r_S}{r}-\frac{r_S}{r_0}}} \,\D u 
 +\frac{-\frac{r_S}{r}\,\EXP{-r/r_S} u^2+1-\frac{r_S}{r_0} }{v
      \sqrt{1-\frac{r_S}{r_0}}  
      + u\sqrt{\frac{r_S}{r}-\frac{r_S}{r_0}} }\,\D v
    \left.\rule{0pt}{7mm}\right\}
\end{equation}
and a similar formula for $c\,\D\tau$. (It is obtained from that for
$\D\ell$ by interchanging $\D u$ and $\D v$, but not $u$ and $v$.)
Taking the limit $r\to r_S$, we find
\begin{equation}
\eqalign{
\D\ell&= r_S \sqrt{1-\frac{r_S}{r_0}}\left(\frac{\D u}{u}
+\frac{\D v}{v}\right) + \frac{r_S}{\text{e} \sqrt{1-\frac{r_S}{r_0}}} \left(u\D u-v\D v\right) \cr
&=r_S \sqrt{1-\frac{r_S}{r_0}}\left(\frac{\D u}{u}
+\frac{\D v}{v}\right) + \frac{1}{2 \sqrt{1-\frac{r_S}{r_0}}}\, \D r
\>,}
\end{equation}
where in the last line we have used $u\D u-v\D v=\frac12 \D (u^2-v^2)$
and (\ref{eq:useful_formulas}).  This is the proper length element of
a freely falling observer who started her fall at $r_0$, with zero
velocity, at the moment when she reaches the horizon. If two observers
fall sufficiently closely after one another, we may use this formula
when one of them is at the horizon.  To measure the spatial distance
between the events of the two horizon crossings, we set $v=u=u_A$ and
$\D v=\D u=u_B-u_A\equiv\Delta u$. Moreover, we obviously have $\D
r=0$.  We then obtain for the spatial and the temporal proper
distances between the two events:
\begin{equation}
\Delta \ell= 2 r_S  \sqrt{1-\frac{r_S}{r_0}} \frac{\Delta u}{u}\>, \quad
c \Delta\tau  = 2 r_S  \sqrt{1-\frac{r_S}{r_0}} \frac{\Delta u}{u}\>.
\end{equation}
This formula, valid only for scenario I, gives finite results for the
distance and time difference as measured by Alice between herself and
her partner. Signs were chosen so that $\D\ell$ is positive for positive $\D r$, hence
the distance to Bob is positive as is Alice's proper time interval,
meaning that Bob falls in \emph{after} her and is positioned towards
the direction from which she came ($\Delta u$ is positive, as
the figure shows). Bob may argue similarly, but for him we would have $\Delta
u=u_A-u_B$ negative.

The result also shows that the proper distance at the horizon between
the two observers becomes larger as $r_0$ increases, which is
reasonable, albeit the $r_0$ dependence can be seen to be weak. More importantly,
the distance increases with increasing $\Delta u$ (at fixed $u$), and we may gather
from the figure that $\Delta u$ becomes larger as the time difference
$t_B-t_A$ between the starting events of Bob's and Alice's journeys
increases. Of course, at some point the validity of the assumption
that both are in a local inertial system, will break down. 
What can still be said is 
that Bob crosses the horizon
after and behind Alice, because their separation is null and she can
send him a signal connecting the crossing events, while he cannot send her
one with the same property.

\section{Conclusions}
Let us briefly summarize the results of our exploration into the
experiences of two adventurers traveling towards the center of a black
hole.  The mathematics of this (frightening) journey of discovery is simple enough
to be presented in classroom and it demonstrates the
utility of alternative coordinate systems in disentangling seem\-ingly
com\-plex situations. The use of simple diagrams adds a visual level to
the com\-pre\-hension of the topic. And of course, the general theme is
exciting, having even made appearance in recent science fiction
movies.

Two basic assumptions were that the black hole is big enough so that
tidal effects are neglible at the horizon and that our protagonists,
Alice and Bob, move by free fall only.  As we have seen, they will not
make any spectacular observations down to the horizon, if they begin
their journey so close to each other that they may be considered still
sharing a common inertial system there. This is a direct
consequence of the equivalence principle and this statement can be
made without any use of coordinates.  KS coordinates are however
useful in discussing how signal exchange between the two observers is
unimpeded by the horizon, although no signal ever crosses it.  That is
possible because the horizon itself locally moves at the speed of light. In
the left panel of figure~1, KS coordinates are handy in visualizing
both signal exchange and the fact that the horizon behaves like an
outgoing signal.

Moreover, KS coordinates show clearly that the horizon crossings of
our two adventurers correspond to different events. In the scenario of
close infallers (scenario I), the proper spatial and temporal
separation that both observers will measure on traversal of the
horizon can be calculated in terms of these coordinates.

This leaves only the second scenario of distant infallers (scenario
II) for the possibility of dramatic observations such as the touching
of Alice's ghostlike image by Bob. Again, KS coordinates do a good job
in studying the situation. In particular, they allow us to calculate
the redshift at the horizon and to compare redshifts of various
observer arrangements. From figure 1 and the redshift formula
(\ref{eq:redshift_6}) in the appendix, we conclude that Bob sees Alice
the more redshifted on crossing the horizon the later he starts his
journey, i.e., the larger their distance when he begins to fall. Also,
she will appear to him at a nonzero distance, when he traverses the
horizon.  Therefore, there are no touching ghosts in this scenario
either.

It might be added that from the point of view of a distant
\emph{stationary} observer, at least scenario II may offer a spectacle
that could be described as ``touching ghosts''. If Alice falls long
before Bob, such a distant observer will see her slowing down near the
horizon, while Bob is still falling fast. Then Bob will also slow down
and both will seem to freeze at the horizon with the coordinate
distance $\Delta r$ between them approaching zero. The idea that this
phemomenon should also find a reflection in local physics may have been
at the origin of the considerations of \cite{augousti12}.  Its
visibility would be brief, because both travelers will redshift out of
the optical frequencies fast. Moreover, it must be emphasized that
\emph{this} touching of ghosts cannot be considered more than an
optical illusion.  There is nothing resembling the touching of ghosts
in the true local physics at the horizon.

Finally, assessing the utility of KS coordinates for our study, they
were useful in the discussion of scenario I and indispensable in the
consideration of scenario II.

\textbf{Acknowledgment} I thank Andrew J. S.  Hamilton from the
University of Color\-ado at Boulder for a very fruitful discussion that
helped me straighten out my ideas about redshift at
the horizon and to finally exorcise the last vestige of a ghost...

\appendix
\section{Redshift calculations}
To derive the redshift $z$ of signals sent by Alice to Bob, we use a
general relativistic expression applicable to the most general cases
\cite{augousti12,narlikar94}
 \begin{equation}
  1+z = \frac{\omega_A^e}{\omega_B^r} 
   = \frac{(U^\alpha V_\alpha)_A}{(U^\alpha V_\alpha)_B}\>.
\label{eq:redshift1}
 \end{equation}
 Herein, the superscript $e$ stands for emission, $r$ for reception,
 and subscripts $A$ and $B$ tag observers (Alice and Bob). $\omega$ is
 the frequency of the signal sent or received.  $U^\alpha$ denotes the
 tangent vector to the null geodesic connecting the emission and
 reception events, to be evaluated at the $A$ or $B$ end,
 re\-spectively, and $V_\alpha$ is Alice's and Bob's four velocity at
 emission and reception, re\-spectively.  We assume $\D\vartheta=
 \D\varphi=0$ for the connecting geodesic, i.e., both observers move
 along purely radial trajectories.

 Let us first have a look at the redshift as long as Alice and Bob are
 both outside the horizon, so we can do the calculation in
 Schwarzschild coordinates. As before, we have the infallers start at
 $r_0$ and follow the same radial path, which means that Bob's
 space\-time trajectory is, once he begins falling, just a time-shifted
 copy of Alice's. The four-velocity components $V^t$ and $V^r$
 are given as a function of $r$ in (\ref{eq:fourvel_schwarz}). 
 Null tangent vectors can be calculated essentially the same way as
 four velocities from a simple metric-induced Lagrangian for
 geodesic motion, using an af\-fine parameter $\lambda$ instead of the
 proper time. More specifics on the method are ex\-hib\-ited below in
 some detail, using KS coordinates. For Schwarzschild coordinates, we
 just give the result:
\begin{equation}
  U^t = c \frac{\D t}{\D\lambda} = \frac{K}{1-\frac{r_S}{r}}\>,\qquad
 U^r = \frac{\D r}{\D\lambda} = K\>,
\end{equation} 
where $K$ is a positive constant that in principle can be determined
from the emission frequency at Alice's end but  is not needed in
the calculation of the ratio (\ref{eq:redshift1}), because it is the
same in the numerator and the denominator.  $U^r=K$ describes an
outgoing light ray; an ingoing ray has $U^r=-K$. The
frequency ratio then evaluates to
\begin{equation}
\hspace*{-1.5cm}\frac{\omega_A^e}{\omega_B^r} =
\frac{g_{tt}(r_A)U^t(r_A) V^t_A + g_{rr}(r_A)U^r(r_A)
 V^r_A}{g_{tt}(r_B)U^t(r_B) V^t_B + g_{rr}(r_B)U^r(r_B) V^r_B} 
= \frac{\sqrt{1-\frac{r_S}{r_0}}-\sqrt{\frac{r_S}{r_B}-\frac{r_S}{r_0}}}{\sqrt{1
-\frac{r_S}{r_0}}-\sqrt{\frac{r_S}{r_A}-\frac{r_S}{r_0}}}\>.
\label{eq:redshift_2}
\end{equation}
Herein, $r_A$ is Alice's position on sending and $r_B$ Bob's position on
receiving the signal. Introducing, as was done in \cite{augousti12},
the local velocity of a falling observer with respect to a stationary
observer at the same position, which is given by 
\begin{equation}
  \tilde v = -\frac{1}{g_{tt}(r)}\frac{\D r}{\D t} 
  = c \frac{\sqrt{\frac{r_S}{r}-\frac{r_S}{r_0}}}{\sqrt{1
      -\frac{r_S}{r_0}}}\>,
\end{equation}
we may reformulate (\ref{eq:redshift_2}) as
\begin{equation}
 \frac{\omega_A^e}{\omega_B^r}= \frac{1-\frac{\tilde v_B}{c}}{1-\frac{\tilde v_A}{c}}\>,
\label{eq:redshift_3}
\end{equation}
a result that by construction holds outside the horizon. The
first thing to note is that if we keep Bob 
stationary, setting $\tilde v_B=0$, the formula reduces to the expression
given in \cite{augousti12} for Alice's redshift as seen by the mother
station. From this we may conclude that the fact that Bob is actually
moving in Alice's direction, will reduce her redshift in comparison
with that seen by a fixed observer. This corresponds to expectations.

Taking the limit $r_0\to\infty$ in (\ref{eq:redshift_2}) to simplify
the formula a bit, we find
\begin{equation}
  \frac{\omega_A^e}{\omega_B^r} =\frac{1-\sqrt{\frac{r_S}{r_B}}}{1
-\sqrt{\frac{r_S}{r_A}}}
\label{eq:redshift_4}
\end{equation}
and this can be immediately compared with the redshift between two
\emph{stationary} observers at $r_A$ and $r_B$, i.e., the momentary
positions of Alice and Bob, respectively. Let us call these Amanda
($A'$) and Brian ($B'$). A light or radio signal sent from Amanda to
Brian will be redshifted, in the Schwarzschild metric, according to
\begin{equation}
  \frac{\omega_{A'}^e}{\omega_{B'}^r} 
  = \frac{\sqrt{1-\frac{r_S}{r_B}}}{\sqrt{1-\frac{r_S}{r_A}}}\>.
\end{equation}
The following inequalities are then evident
\begin{equation}
  \frac{\omega_{A'}^e}{\omega_{B'}^r} = 
  \left[\frac{\left(1-\sqrt{\frac{r_S}{r_B}}\right)\left(1
        +\sqrt{\frac{r_S}{r_B}}\right)}{\left(1-\sqrt{\frac{r_S}{r_A}}\right)\left(1
        +\sqrt{\frac{r_S}{r_A}}\right)}\right]^{1/2} < \sqrt{\frac{\omega_A^e}{\omega_B^r}} 
  < \frac{\omega_A^e}{\omega_B^r}\>,
\end{equation}
because $r_A<r_B$ implies that $\left(1
  +\sqrt{\frac{r_S}{r_B}}\right)/\left(1
  +\sqrt{\frac{r_S}{r_A}}\right) < 1$ and, due to
$\omega_A^e/\omega_B^r>1$, the square root of the ratio is smaller
than the ratio itself. The result means that the redshift between
Alice and Bob is \emph{larger} than the one between two stationary
observers at their momentary positions on emission and reception
of the signal.  This also corresponds to our intuition, as Alice is
ahead and always falling at a larger local velocity than Bob, so on
top of the gravitational redshift there should be a redshift by the
Doppler effect. Moreover, in a Newtonian universe, the redshift
between Alice and Bob should increase during the fall, as she always
experiences a stronger acceleration.  However, the world is not
Newtonian and our intuition may not be too helpful in assessing what
happens near the horizon, because the local velocities of both
travelers approach $c$ there.  Formulas (\ref{eq:redshift_2}) and
(\ref{eq:redshift_3}) get indefinite.  The idea that the ratio
approaches one at the horizon and therefore Bob will see Alice without
redshift when he crosses the horizon, is too simplistic. If it were
true, we would really have a ghostly image of Alice that reappears
back in the optical spectrum after having been redshifted out of it
before. This would still not be a \emph{touching} ghost but at least
an interesting optical effect.  Alas, it is not so.

Clearly, our analytical redshift formulas are incomplete in that they
do not give us Alice's redshift as observed by Bob as a function of
her or his proper time. To obtain that, we would need to have Bob's
position on signal reception for each of Alice's emission positions,
i.e.  $r_B(r_A(\tau))$, which could be calculated, in principle, by
intersecting the world line of Bob with Alice's future light cone.
This is not analytically trivial, to say the least. Once we had
$r_B(r_A)$, we might take the limit $r_A\to r_S$ in
(\ref{eq:redshift_2}), using de l'Hospital's rule. Without this
information, our Schwarzschild coordinate expressions are not useful
in performing the limit.

It is here, where the power of KS coordinates shows up again,
permitting an additional step. 
To evaluate the redshift formula, we first calculate
the null geodesics of radial light signals in KS coordinates. After dropping the angular
coordinates (which are constant), the appropriate Lagrangian is given by
 \begin{equation}
 L =\frac{4 r_S^3}{r} \, \EXP{-r/r_S}\left(\mathring v^2-\mathring u^2\right)
 \label{eq:lagrangian}
 \end{equation}
 where $\mathring v = \D v/\D\lambda$, $\mathring u = \D u/\D\lambda$
 and $\lambda$ is an affine parameter again.  The Lagrangian equation
 of motion for $v$ then reads
 \begin{equation}
  \hspace*{-1.5cm} \frac{\D}{\D\lambda} \pabl{L}{\mathring v} -
 \pabl{L}{v} =\frac{\D}{\D\lambda} \left( \frac{8 r_S^3}{r} \,
  \EXP{-r/r_S} \mathring v\right) 
 - \pabl{}{v} \left(\frac{4 r_S^3}{r} \, \EXP{-r/r_S}\right)
  \left[\mathring v^2-\mathring u^2\right] = 0\>.
 \label{eq:eq_for_v}
 \end{equation}
 The second equation of motion is, as usual, more easily obtained
 using the Lagrangian itself, which is zero here (null geodesics!), so
 that (\ref{eq:lagrangian}) gives us:
 \begin{equation}
 \mathring v^2 = \mathring u^2\>,
 \end{equation}
 which for an outgoing light ray implies $\mathring v= \mathring u$. But then the second term in (\ref{eq:eq_for_v}) is zero and we find
 \begin{equation}
 \frac{8 r_S^3}{r} \, \EXP{-r/r_S} \mathring v = 2 D^{-1} \mathring v = \text{const.}
 \label{eq:const_mot}
 \end{equation}
Hence, we may write
 \begin{equation}
U^v=\mathring v =  D K\>,\quad U^u= \mathring u =  D K
 \end{equation}
with some constant $ K$. At the horizon, $D=\frac{\text{e}}{4 r_S^2}$, therefore $ U^v=U^u = \text{const.}$
The four velocity of an infalling observer in KS coordinates was calculated in (\ref{eq:vdot_udot}). We find
 \begin{equation}
\eqalign{g_{vv}(r) &U^v V^v + g_{uu}(r) U^u V^u = \frac1D \, D K\left(V^v-V^u\right) \cr
&=  K \frac{c}{2r_S} \frac{1}{1-\frac{r_S}{ r}} (u-v) \left(\sqrt{1-\frac{r_S}{r_0}}+\sqrt{\frac{r_S}{r}-\frac{r_S}{r_0}}\right)
\cr
&= \frac{ K c}{2 r_S} \frac{r}{r_S}\EXP{r/r_S} \frac{\sqrt{1-\frac{r_S}{r_0}}+\sqrt{\frac{r_S}{r}-\frac{r_S}{r_0}}}{u+v}
}
 \end{equation}
and this leads to a redshift formula,
 \begin{equation}
   \frac{\omega_A^e}{\omega_B^r} =\frac{u_B+v_B}{u_A+v_A} \frac{r_A}{r_B} \EXP{(r_A-r_B)/r_S} 
 \frac{\sqrt{1-\frac{r_S}{r_0}} +\sqrt{\frac{r_S}{r_A}-\frac{r_S}{r_0}}}{\sqrt{1
       -\frac{r_S}{r_0}}+\sqrt{\frac{r_S}{r_B}-\frac{r_S}{r_0}}}\>,
\label{eq:redshift_5}
 \end{equation}
 that is manifestly regular at the horizon.  Surprising as it may
 seem, (\ref{eq:redshift_2}) and (\ref{eq:redshift_5}) are the same
 result; a small calculation benefitting from  the simple
 light cone re\-pre\-sen\-tation in KS coordinates shows
 (\ref{eq:redshift_2}) to follow from (\ref{eq:redshift_5}) (and vice
 versa).

 By taking the limit $r_A, r_B\to r_S$, we get Alice's redshift as
 seen by Bob on crossing the horizon:
 \begin{equation}
   1+z= \frac{\omega_A^e}{\omega_B^r}\Big\vert_{r_A=r_B=r_S} =\frac{u_B+v_B}{u_A+v_A}  =\frac{v_B}{v_A} = \frac{u_B}{u_A}\>.
\label{eq:redshift_6}
 \end{equation}
 This is finite and $z$ is different from zero. In fact, comparing the
 pictures from scenario I and scenario II, we see that the ratio
 $\frac{v_B}{v_A}$ of the points where the trajectories cross the
 horizon becomes the larger, the later Bob starts his journey. If he
 starts very much later than Alice, he will not see her at all on
 crossing the horizon, because her image will have shifted out of the
 optical spectrum. Since the quantity $u+v$ is constant along ingoing
 light rays, we may also infer from the figure that at least in the
 right panel the ratio $\frac{u_B+v_B}{u_A+v_A}$ increases with
 Alice's proper time for events on Alice's and Bob's trajectories that
 are connected by an outgoing light ray. (The length of outgoing rays
 for Alice's signals \sfsl{1} through \sfsl{5} increases with their
 ordinal number, faster than $u_A +v_A$ does.)

 Outside the horizon, the factor $ \frac{r_A}{r_B}
 \EXP{(r_A-r_B)/r_S}$ is smaller than one and may first decrease but
 approaches one close to the horizon. The last factor  of (\ref{eq:redshift_5}) increases as
 $r_A$ decreases, and also approaches one near the horizon, where
 everything is dominated by the first factor.

 Inside the horizon, $r_B$ is \emph{smaller} than $r_A$ for events on
 the two trajectories con\-nected by an outgoing light ray, as may be
 seen easily by (mentally) constructing a hyperbola
 $v^2-u^2=\text{const.}$ through the endpoint of Alice's signal
 \sfsl{4} in the left or \sfsl{5} in the right panel. Hence all three
 leading factors of (\ref{eq:redshift_5}) grow during the fall towards
 the singularity and this cannot be compensated by the last factor
 that is decreasing more slowly.  Note that the formula seems to
 predict infinite blueshift as Alice approaches the singularity
 ($r_A\to 0$).  However, $r_B$ becomes zero \emph{first} (at signal
 \sfsl{5} in the left panel of the figure), and there is, along all of
 Alice's and Bob's trajectories, no pair $(r_A,r_B)$ connected by an
 outgoing light ray that satisfies $r_A=0$. Rather, it is Alice's
 \emph{redshift} that becomes infinite for Bob the moment \emph{he} hits the
 singularity.

 So the overall result seems to be that Alice's signals arrive at
 Bob's positions at consecutive (proper) times with monotonously increasing
 redshift, outside the horizon as well as inside. What has been shown
 rigorously here is that there is a nonzero redshift at the horizon
 and that it increases with Bob's time delay in following Alice. Inside the
 horizon, there can be little doubt, given the structure of the
 redshift expression, that the redshift continues to grow  towards
 the singularity. Outside the horizon, the situation is also clear far
 away from it, where Newtonian approximations apply, indicating that
 the redshift between Alice and Bob must increase as a function of
 time. Finally, there is no good reason to expect non-monotonous behavior
 at intermediate distances to the horizon.

\rule{0pt}{0pt} \\

\bibliographystyle{unsrt}

\end{document}